\documentclass[prl, twocolumn]{revtex4}
\usepackage{amsmath}
\usepackage{amssymb}
\usepackage{graphicx}
\usepackage{subfigure}
\usepackage{verbatim}
\usepackage{hyperref}
\usepackage{xfrac}

\makeatletter
\newcommand*{\rom}[1]{\expandafter\@slowromancap\romannumeral #1@}
\makeatother

\begin{document}
\title{Umbilic Lines in Orientational Order}
\author{Thomas Machon}
\author{Gareth P. Alexander}
\affiliation{Department of Physics and Centre for Complexity Science, University of Warwick, Coventry CV4 7AL, United Kingdom.}
\begin{abstract}
Three-dimensional orientational order in systems whose ground states possess non-zero, chiral gradients typically exhibits line-like structures or defects: $\lambda$ lines in cholesterics or Skyrmion tubes in ferromagnets for example. Here we show that such lines can be identified as a set of natural geometric singularities in a unit vector field, the generalisation of the umbilic points of a surface. We characterise these lines in terms of the natural vector bundles that the order defines and show that they give a way to localise and identify Skyrmion distortions in chiral materials -- in particular that they supply a natural representative of the Poincar\'{e} dual of the cocycle describing the topology. Their global structure leads to the definition of a self-linking number and helicity integral which relates the linking of umbilic lines to the Hopf invariant of the texture. 
\end{abstract}
\maketitle

Three-dimensional systems described by orientational order, and more generally vector fields, are often seen to contain line-like geometrical features. For example, in cholesteric liquid crystals as $\lambda$ lines~\cite{bouligand74,bouligand78,kleman69,smalyukh02,ackerman14}, or dislocations, and as double twist cylinders in blue phases~\cite{meiboom81,wright89}, in superfluids they arise as the cores of vortices~\cite{mermin76,volovik76,anderson77,walmsley12}, and they can be identified in the Skyrmion~\cite{note terminology} textures of chiral ferromagnets~\cite{rossler06,muhlbauer09,yu10,yu12,milde13} and Bose-Einstein condensates~\cite{khawaja01,leslie09,choi12,choi13,ku14,donadello14}. These structures are being manipulated with increasing precision in experimental systems, providing a variety of topologically non-trivial textures in ordered media with ample potential for the development of structured metamaterials~\cite{lavrentovich11,lavrentovich13,musevic11}, fluidic photonics~\cite{musevic13} and ultralow current spintronics~\cite{kiselev11,duine13,fert13}. Skyrmions in chiral magnets are often described as magnetic whirls that encode a topological winding. The same topological profiles can be imprinted on spin configurations of Bose-Einstein condensates, superfluids and electronic states. Skyrmions also arise in liquid crystals~\cite{smalyukh10,chen13,fukuda10,ackerman12b,ackerman14,pandey14}, where the nature of nematic order allows for a greater variety of three-dimensional textures including knots, a focus of activity across several disciplines~\cite{dennis10,tkalec11,kleckner13,machon13,martinez14,tasinkevych14}, which can then be entangled by Skyrmions~\cite{machon14}. All of these display a blend of topological features, giving robustness, with a geometrical character that comes from energy minimisation. In this paper we show that these objects can be described as a set of natural geometric singularities in a vector field, which we call umbilic lines, whose topological significance comes through their representation of the Poincar\'e dual to the Euler class of a naturally defined vector bundle. 

The umbilics we describe are generic traits, exhibited by any vector or line field, and closely analogous to several other structural degeneracies, for instance the lines of circular polarisation (C lines) in a generic electromagnetic field~\cite{nye83,nye87,hajnal87a,hajnal87b} and the umbilic points of surfaces~\cite{Hilbert,berry77} and random fields~\cite{berry00,berry01,dennis02,flossmann05,flossmann08,vitelli09,beuman12,beuman13}, from which the terminology derives. Indeed, degeneracies in geometrical physical properties are commonplace; their relevance and significance for the description of global or structural aspects of physical behaviour, and the identification of generic traits, has been cemented since Berry's influential paper on the geometrical phase in quantum mechanics~\cite{berry84,berry89}. These degeneracies are not always random; for instance, it has been known for a long time that the clear blue sky is polarised and that there are points of degeneracy (or singularity) in that polarisation, the Babinet and Brewster points either side of the sun, and the Arago and second Brewster points either side of the anti-sun. Away from these points the polarisation is elliptical and the degeneracy is to a state of circular polarisation, about which the principal axes of the polarisation ellipse rotate by $\pi$, corresponding to a $+1/2$ index. That there are four such points is a reflection of the fact that the sky has the topology of a sphere and an illustration of the Hopf index theorem~\cite{berry04c,Hopf}. The same Berry phases control aspects of the response in magnetic Skyrmions~\cite{neubauer09,everschor11,schultz12}. 

We give general local and global descriptions of orientational order in three dimensions, identifying umbilic lines as a key natural feature characterising the ordered material. The flux of umbilics through any surface is identified with the Skyrmion charge through an application of the Gauss-Bonnet-Chern theorem. The umbilics convey the topology in the orientational order since they are Poincar\'e dual to twice the Euler class of the two-plane bundle of directions that are everywhere orthogonal to the local director. The extended nature of umbilics entails additional global properties that we describe in terms of self-linking numbers. An analogue of helicity connects to the Hopf invariant in a manner parallel to that for fluid flows~\cite{moffatt69,moffatt92,arnold86}. Of additional note is a connection of our local description with recent work of Efrati and Irvine~\cite{efrati14} on chiral shapes and structures. In their work they introduced a chirality pseudotensor to characterise the direction of twisting and sense of chirality of various materials. We demonstrate a general connection with the shape operator for a surface or vector field, from which their chirality pseudotensor can be derived, and show that natural benefits are gained by considering both operators. Analogous geometrical structures to the umbilics we describe have been discussed by Thorndike {\sl et al.}~\cite{thorndike78}, \v{C}opar {\sl et al.}~\cite{copar13}, and Beller {\sl et al.}~\cite{beller14}, although not with the topological aspect that we give here. 

Umbilic lines acquire extra significance in chiral materials where the ground state has non-zero gradients. In these materials the vortex configurations surrounding umbilics correspond to local minima of the free energy. For concreteness, then, we consider ordered materials described by a unit magnitude vector field ${\bf n}$, called the director, and whose free energy has a form similar to
\begin{equation}
F = \frac{K}{2} \int_{\mathbb{R}^3} \Bigl[ |\nabla \mathbf{n}|^2 + 2 q_0 \mathbf{n} \cdot \nabla \times \mathbf{n} \Bigr] d^3x ,
\label{eq:f}
\end{equation}
where $K$ is an elastic constant and $q_0$ is the strength of the chirality, or magnitude of spin-orbit coupling. This is the Frank free energy~\cite{deGennesProst} for a bulk cholesteric under the one elastic constant approximation and also the Hamiltonian for elastic distortions in a chiral ferromagnet with a Dzyaloshinskii-Moriya interaction~\cite{dzyaloshinskii58,moriya60}.

\section{Local Geometry of Orientational Order}

The local geometry of the orientational order at any point in the material can be given the following generic characterisation. Since ${\bf n}$ is a unit vector its gradients are always perpendicular to itself. They separate naturally into gradients along the direction of the local orientation, known as bend distortions, and gradients along orthogonal directions, which correspond to splay, twist, and saddle-splay deformations. These latter represent a linear transformation on the two-dimensional plane of directions perpendicular to the local orientation, which is how we shall think of them. If ${\bf d}_1, {\bf d}_2$ are unit vectors orthogonal to ${\bf n}$, such that $\{{\bf n},{\bf d}_1,{\bf d}_2\}$ form a right-handed set, then the orthogonal gradients can be expressed in this local basis as a $2\!\times\!2$ real matrix with the decomposition 
\begin{equation}
\begin{split}
\nabla_{\perp}{\bf n} & = \frac{\nabla\cdot{\bf n}}{2} \begin{bmatrix} 1 & 0 \\ 0 & 1 \end{bmatrix} + \frac{{\bf n}\cdot\nabla\times{\bf n}}{2} \begin{bmatrix} 0 & -1 \\ 1 & 0 \end{bmatrix} \\
& \qquad + \begin{bmatrix} \Delta_1 & \Delta_2 \\ \Delta_2 & -\Delta_1 \end{bmatrix} .
\end{split}
\label{eq:local_geom}
\end{equation} 
The first component is an identity transformation on the orthogonal directions, while the second is a skew transformation that provides the orthogonal directions with a complex structure. The coefficients of these two terms are known as the mean curvature and the mean torsion~\cite{rogers12,Aminov}, respectively, and correspond to scalar and pseudoscalar parts of the transformation. The last part is a deviatoric, or spin 2, component, $\Delta$, that conveys the local `shear'. Its eigenvectors can be used to define principal directions of the curvature of the local orientation. In the case where ${\bf n}$ is the normal to a surface, the transformation \eqref{eq:local_geom} is the derivative of the Gauss map, or shape operator of the surface, and the principal directions alluded to are the directions of principal curvature. 
 
The geometric character of the orientational order is conveyed by this generic local description. For instance the twist, ${\bf n}\cdot\nabla\times{\bf n}$, encodes the chirality of the order with surfaces of zero twist separating regions of different handedness. Zeros of the deviatoric part of $\nabla_{\perp}{\bf n}$ occur along lines in three dimensions -- they require both $\Delta_1$ and $\Delta_2$ to vanish and hence are codimension two -- and correspond to degeneracies of the principal curvatures, or singularities of the principal curvature directions. They are the umbilics of the orientational order. At such points the transverse gradients are isotropic meaning that umbilics are naturally associated with the centres of vortex-like structures. 
A basis-independent expression can be given for these zeros purely in terms of the director; they correspond to the vanishing of the non-negative quantity 
\begin{equation}
\begin{split}
|\Delta|^2 & = \biggl( \frac{\nabla\cdot{\bf n}}{2} \biggr)^2 + \biggl( \frac{{\bf n}\cdot\nabla\times{\bf n}}{2} \biggr)^2 \\
& \quad - \frac{1}{2} \nabla \cdot \Bigl[ {\bf n} \bigl( \nabla\cdot{\bf n} \bigr) - \bigl( {\bf n}\cdot\nabla \bigr) {\bf n} \Bigr] ,
\end{split}
\end{equation}
this being the 2-norm of the matrix $\Delta$. In the vicinity of an umbilic we may write 
\begin{equation}
\Delta = |\Delta| \begin{bmatrix} \cos\theta & \sin\theta \\ \sin\theta & - \cos\theta \end{bmatrix} ,
\label{eq:delta_matrix}
\end{equation}
and around the umbilic the angle $\theta$ winds by an integer multiple of $2\pi$. An umbilic is said to be generic if it has a winding number of $\pm 1$. Umbilics with higher winding do occur, and indeed those with winding $+2$ are often observed in vortices, double twist cylinders and Skyrmions, with the degeneracy coming from an imposed axisymmetry. This can persist in certain experimental settings, either where there is axisymmetry or hexagonal symmetry, however, from a structural point of view they can be seen as a composite of winding $\pm 1$ umbilics, into which they break apart under perturbation. The generic umbilic points of a surface have been classified as stars, lemons, or monstars~\cite{berry77}, according to the local profile that the principal curvature directions have in the vicinity of the umbilic point. The same local structure arises in the rotation of the eigenvectors of $\Delta$ around the umbilic lines of a generic vector field in three dimensions -- it is not hard to see that the eigenvectors of $\Delta$ rotate by only half as much as $\Delta$ itself, {\it i.e.} by a half-integer multiple of $2\pi$, this being the familiar phenomenon of a real Berry phase~\cite{berry84,berry89} -- so that the local profile of an umbilic line may be classified in the same way as the umbilic points of a surface or the C lines of generic electromagnetic fields~\cite{nye83}. As this description is classical~\cite{berry77,thorndike78} we choose to focus on complementary aspects, making more use of the language of vector bundles~\cite{MilnorStasheff}.  

\begin{figure}
\includegraphics{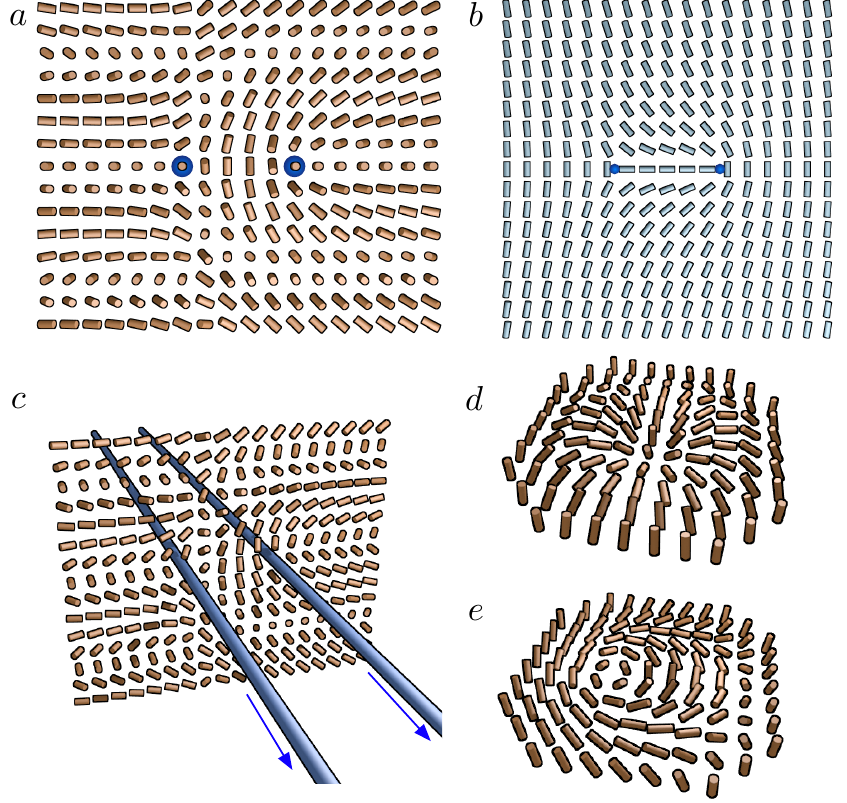}
\caption{Umbilic lines. (a) $\lambda^+$ $\lambda^-$ dislocation in a cholesteric. It contains two generic umbilics. This combination of two generic umbilics can be identified as half a Skyrmion, or a meron. (b) Eigenvector field of $\Pi$ for the dislocation. The two umbilics have opposite windings of the eigenvectors, but are counted with the same sign as $\mathbf{n}$ rotates by $\pi$ when passing from one to the other. (c) Umbilic lines can be oriented in three dimensions. (d) \& (e) Typical centres of vortices, based on curvature and torsion distortions respectively, that correspond to winding number $2$ non-generic umbilics.}
\label{fig:vort}
\end{figure}

Illustrations can be given for a number of simple field conformations corresponding to textures observed in superfluid vortices, magnetic Skyrmions and the $\lambda$ lines of cholesteric liquid crystals, shown in Fig.~\ref{fig:vort}. Two-dimensional sections of the full textures are shown on surfaces intersecting the umbilics transversally. The eigenvectors of $\Pi$ shown in (b) identify the profiles of the umbilics in (a). Away from the umbilics this eigendirection coincides with the pitch axis of the helical texture. 

We explain later that the windings just described are equivalent to the indices of the umbilics up to a sign, which arises because the orientation of ${\bf n}$ reverses from one umbilic to another. (The discrepancy is therefore absent if ${\bf n}$ is the normal to a surface, since then there is no reversal of its orientation between umbilics, and the index is simply the local winding of the eigenvectors of $\Delta$.) The sum of the indices of the umbilics yields a topological invariant, equal to four times the Skyrmion number of the texture on the two-dimensional surface. 
This calculation of the Skyrmion number is dual to the usual method of integrating a curvature form~\cite{belavin75,MantonSutcliffe} 
\begin{equation}
q = \frac{1}{8\pi} \int_{\Sigma} \epsilon_{abc} n^a \partial_i n^b \partial_j n^c\, dx^i \wedge dx^j ,
\end{equation}
which computes the degree of the map ${\bf n}:\Sigma\to S^2$. The duality can be viewed as entirely analogous to that in the classical differential geometry of surfaces between the Gauss-Bonnet theorem -- integral of the Gaussian curvature -- and the calculation of the Euler characteristic using umbilic points of the surface.  

Two additional aspects of the local description are worth bringing out before giving a global treatment. First, the complex structure $J=\bigl[\begin{smallmatrix} 0 & -1 \\ 1 & 0 \end{smallmatrix}\bigr]$ acts on the shape operator~\eqref{eq:local_geom} by simple composition to produce an equivalent linear transformation 
\begin{equation}
\begin{split}
\chi = J\circ\nabla_{\perp}{\bf n} & = - \frac{{\bf n}\cdot\nabla\times{\bf n}}{2} \begin{bmatrix} 1 & 0 \\ 0 & 1 \end{bmatrix} + \frac{\nabla\cdot{\bf n}}{2} \begin{bmatrix} 0 & -1 \\ 1 & 0 \end{bmatrix} \\
& \qquad + \begin{bmatrix} -\Delta_2 & \Delta_1 \\ \Delta_1 & \Delta_2 \end{bmatrix} .
\end{split}
\label{eq:local_chi}
\end{equation}
This transformation, although equivalent to the shape operator, is independent from it in the sense that they are orthogonal with respect to the natural inner product between matrices, $\textrm{Tr}(\chi^T\nabla_{\perp}{\bf n})=0$, a property that turns out to be of some importance in the analysis of the topology of umbilics. $\chi$ is the chirality pseudotensor introducted recently by Efrati and Irvine~\cite{efrati14} in their study of chiral materials; our description of it provides a complementary perspective to theirs. The spin 2 component $\Pi=J\circ\Delta$ carries equivalent information to $\Delta$. Its eigenvectors can be thought of as defining directions of principal torsion (or twist) and are simply rotated relative to those of $\Delta$ by $\pi/4$ about the director. Nonetheless, for materials with chiral interactions or structure, it is these directions, rather than the eigenvectors of $\Delta$, that are often more visibly recognisable, as is elegantly illustrated in the experiments of Armon {\sl et al.}~\cite{armon11} on the opening of chiral seed pods, and those of Efrati and Irvine~\cite{efrati14} on stretched elastic sheets. 
For this reason, we plot the eigenvectors of $\Pi$ rather than those of $\Delta$ in all figures in this paper. 

Second, if the umbilic forms a closed loop then the winding of the angle $\theta$ along the contour length of the umbilic (a longitude) conveys a second integer invariant. Of course, this latter depends on both the choice of longitude and the homotopy class of the local trivialisation $\{{\bf d}_1,{\bf d}_2\}$. We describe this in detail later.

\section{Global Definition of Umbilics and the Topology of Vector Fields}

The umbilic lines we have identified convey topological information about the vector field ${\bf n}$, specifically the Euler class of the orthogonal 2-plane bundle, $\xi$, through a combination of the Gauss-Bonnet-Chern theorem~\cite{chern44,chern45,chern46,dirac31} and Poincar\'e duality. The curvature of a vector bundle represents a characteristic cohomology class that can be integrated over homology cycles to produce topological invariants (Euler numbers). By Poincar\'e duality this can be viewed as a homology class and the umbilics furnish a representative of this Poincar\'e dual. We shall develop this general topology of orientational order in terms of a natural connection on the vector bundle that the spin 2 transformation $\Delta$ takes values in. 

\begin{figure}
\begin{center}
\includegraphics{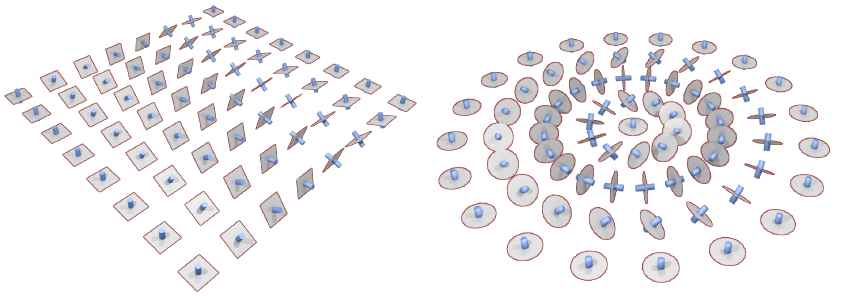}
\end{center}
\caption{The splitting $T\mathbb{R}^3 \cong L_\mathbf{n} \oplus \xi$ induced by orientational order. {\em Left:} A simple helical director profile, corresponding to the ground state of a cholesteric liquid crystal or helimagnet. {\em Right:} An idealised, axisymmetric Skyrmion profile, the central portion of which is a standard idealised model of double twist cylinders in blue phases.}
\label{fig:2plane}
\end{figure}

As before, let $\mathbf{n}$ be a unit vector field in $\mathbb{R}^3$; then it defines a splitting of the tangent bundle 
\begin{equation}
T\mathbb{R}^3 \cong L_\mathbf{n} \oplus \xi,
\label{eq:tangsplit}
\end{equation}
into a real line bundle, $L_\mathbf{n}$, of vectors parallel to $\mathbf{n}$ and a rank 2 real vector bundle, $\xi$, of vectors orthogonal to $\mathbf{n}$. Examples of this splitting are shown in Fig.~\ref{fig:2plane}. The gradients of ${\bf n}$ take values in $T\mathbb{R}^{3\,*}\otimes\xi\cong(L_{{\bf n}}^*\otimes\xi)\oplus(\xi^*\otimes\xi)$, with the first part containing the bend deformations and the second the perpendicular gradients given locally in \eqref{eq:local_geom}. By complete analogy with the classical differential geometry of surfaces we call these perpendicular gradients the shape operator of the orientational order and think of it as a linear transformation on the orthogonal 2-planes. 

The shape operator decomposes in the following way. Local rotations about the director field correspond to a local symmetry and endow the bundle $\xi$ with a natural $SO(2)$ action, under which the tensor product $\xi^*\otimes\xi$ splits into a direct sum $\mathcal{I} \oplus \mathcal{J} \oplus E$, where $\mathcal{I}$ and $\mathcal{J}$ are trivial line bundles and $E$ is a rank 2 vector bundle. Correspondingly, the shape operator has the decomposition  
\begin{equation}
\nabla_{\perp}{\bf n} = \frac{\nabla \cdot \mathbf{n}}{2} \mathbb{I}_\xi + \frac{\mathbf{n} \cdot \nabla \times \mathbf{n}}{2} J + \Delta,
\label{eq:decomp}
\end{equation}
given previously in \eqref{eq:local_geom}. Here, $\mathbb{I}_{\xi}$ is the identity transformation and $J={\bf n}\times\,$ is a complex structure on $\xi$, now defined globally, while $\Delta$ has the global definition 
\begin{equation}
\Delta({\bf v}) = \frac{1}{2} \Bigl[ \bigl( {\bf v}\cdot\nabla \bigr) {\bf n} + {\bf n} \times \bigl( {\bf n}\times{\bf v} \cdot \nabla \bigr) {\bf n} \Bigr] ,
\end{equation} 
for any ${\bf v}$ in $\xi$. The umbilics are identified with the zeros of $\Delta$; they encode the topology of the orientational order. 

Without repeating too much of the local description given previously, we record simply that composition with the complex structure again yields the chirality pseudotensor~\cite{efrati14}, $\chi=J\circ\nabla_{\perp}{\bf n}$, and that the spin 2 component of this, $\Pi=J\circ\Delta$, can be expressed globally as 
\begin{equation}
\Pi({\bf v}) = \frac{1}{2} \Bigl[ {\bf n}\times \bigl( {\bf v}\cdot\nabla \bigr) {\bf n} - \bigl( {\bf n}\times{\bf v} \bigr) \cdot \nabla {\bf n} \Bigr] ,
\label{eq:def_Pi}
\end{equation}
for any ${\bf v}$ in $\xi$, and is also a section of $E$. We reiterate that although $\Pi$ and $\Delta$ may be obtained one from the other by composition with the complex structure, they are orthogonal with respect to the natural inner product $\langle \Pi,\Delta \rangle = \tfrac{1}{2} \textrm{Tr}(\Pi^T\Delta)$. At the same time, they share the same magnitude and the same zero set, the umbilics ${\cal U}$. Thus, away from umbilic lines, $\Pi$ and $\Delta$ represent two non-zero orthogonal sections of $E$, and so they provide a basis for $E|_{\mathbb{R}^3-\mathcal{U}}$ and also define a natural connection 1-form 
\begin{equation}
A = \frac{\langle \Pi,\nabla \Delta \rangle}{\langle \Delta, \Delta \rangle} .
\label{eq:def_A}
\end{equation}
This 1-form conveys the fundamental topology of the orientational order. 

We give in \eqref{eq:def_A} a global definition of $A$ that emphasises its naturality and makes clear how it arises. In a local chart it can be represented in terms of the eigenvectors, $\mathbf{p}^{+}$ and $\mathbf{p}^{-}$, of $\Delta$ and has components $A_i=2p^{-}_{a} \partial_i p^{+}_{a}$, an expression perhaps closer in form to the usual way of writing Berry connections. However, one must keep in mind that this representation is only local in the present context; otherwise it only serves to obfuscate the natural structure. The exterior derivative of $A$, $\Omega_E=dA$, is the (Berry) curvature of the vector bundle $E$. It is equal to twice the curvature of $\xi$ on account of $E$ being a spin 2 subbundle of $\xi^*\otimes\xi$. (This is also evident from the local expression in terms of the eigenvectors ${\bf p}^{\pm}$.)

\begin{figure}
\begin{center}
\includegraphics{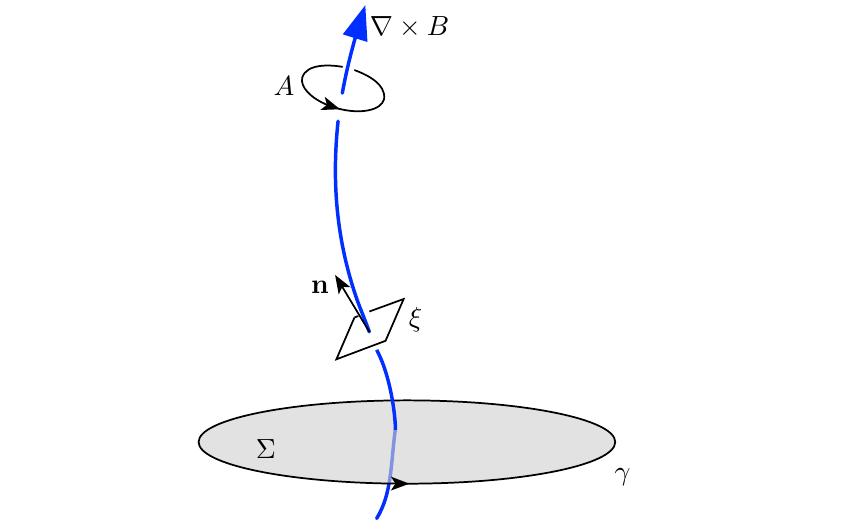}
\end{center}
\caption{Anatomy of an umbilic line. \textit{Top}. The vector $\nabla \times B$ is tangent to generic umbilic lines, giving them a natural orientation. This orientation is compatible with that induced by $A$. \textit{Middle}. The vector field $\mathbf{n}$ and associated orthogonal plane $\xi$. Note that in general the tangent vector of the umbilic does not align with $\mathbf{n}$. $\xi$ is oriented by $\mathbf{n}$ and so gives a natural cycle along which to measure the winding. For a generic umbilic if $\mathbf{n} \cdot \nabla \times B>0$ then the eigenvectors of $\Delta$ have a $+1/2$ profile and if $\mathbf{n} \cdot \nabla \times B<0$ the eigenvectors have a $-1/2$ profile. \textit{Bottom}. A measuring cycle $\gamma$ bounding a surface $\Sigma$. Whether $\gamma$ measures the winding of the umbilic as positive or negative depends on whether the umbilic -- oriented by $\nabla \times B$ -- intersects $\Sigma$ positively or negatively.}
\label{fig:anat}
\end{figure}

A fundamental property of an umbilic is given by the integral of the connection 1-form \eqref{eq:def_A} around any simple closed loop $\gamma$ encircling the umbilic. This is the holonomy of the connection, or Berry phase. Let $\Sigma$ be an oriented surface with boundary $\gamma$, as in Fig.~\ref{fig:anat}. If $\Sigma$ is generic then the umbilics will intersect it transversally in some number of distinct points. Let $\{D_i\}$ be disjoint small discs in $\Sigma$ about each of these intersection points and denote their boundaries by $\{C_i\}$. Then $|\Delta|$ is nowhere zero on $\Sigma-\{D_i\}$ and $\partial(\Sigma-\{D_i\})=\gamma\sqcup\{-C_i\}$ so that by Stokes' theorem 
\begin{equation}
\int_{\gamma} A - \sum_{i} \int_{C_i} A = \int_{\Sigma-\{D_i\}} dA = \int_{\Sigma-\{D_i\}} \Omega_E . 
\end{equation}
To extend the analysis over the umbilics, we can introduce a local trivialisation of $L_{{\bf n}}\oplus\xi$ on each of the $D_i$, which we denote by the orthonormal basis $\{\mathbf{n}, \mathbf{d}_1, \mathbf{d}_2 \}$. Then $\mathbf{e}_1=\mathbf{d}_1 \otimes \mathbf{d}_1 - \mathbf{d}_2 \otimes \mathbf{d}_2$ and $\mathbf{e}_2=\mathbf{d}_1 \otimes \mathbf{d}_2 + \mathbf{d}_2 \otimes \mathbf{d}_1$ form a local basis for $E$. In such a basis $\Delta= |\Delta| \cos \theta\, \mathbf{e}_1 + |\Delta| \sin \theta\, \mathbf{e}_2$ and $A=d \theta + 2 \omega$ where $\omega_i=(\mathbf{d}_{2})_{a} \partial_i (\mathbf{d}_{1})_{a}$ is the connection 1-form on $\xi$ for this basis (the factor of two accounts for $E$ being a spin 2 bundle). With this, one finds at once the Gauss-Bonnet-Chern theorem in the form 
\begin{align}
\frac{-1}{2\pi} \int_{\Sigma} \Omega_E + \frac{1}{2\pi} \int_{\gamma} A & = \sum_{i} \frac{1}{2\pi} \int_{C_i} d\theta , \notag \\
& = \sum_{i} \textrm{index}_\Sigma \; U_i ,
\label{eq:gbc}
\end{align} 
which defines the index of an umbilic, an integer associated to its intersection with a surface. The sum of these over all of the umbilics is an Euler number of the bundle $E$. 
Note that, as alluded to earlier, the index of an umbilic is not simply the winding number of the angle $\theta$ that characterises the local profile as that is measured relative to a local basis for $\xi$, while the index uses the orientation on the surface $\Sigma$; there is, therefore, an additional sign according to whether $d\theta$ is cooriented with $C_i$ or not. 

It is instructive to contrast this situation, where the type of profile does not carry topological information, with that of umbilical points of surfaces where the type of profile does carry topological information about the sign of the winding. This occurs because in the case of surfaces $\mathbf{n}$ is fixed to be the normal to $\Sigma$, and so $d \theta$ is always cooriented with $C_i$.

\begin{figure}
\includegraphics{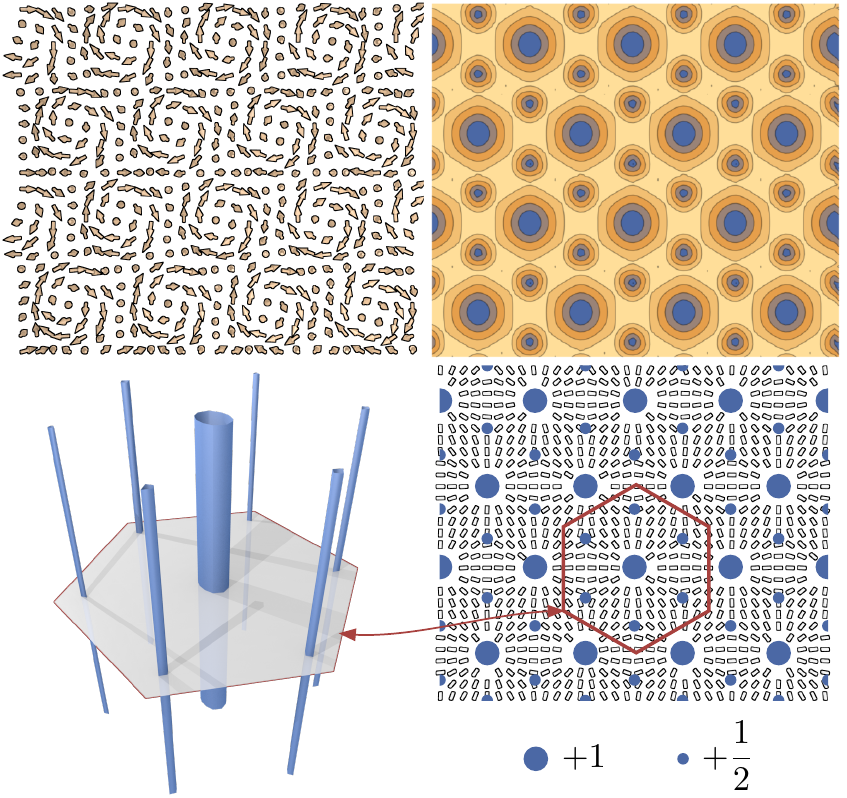}
\caption{Skyrmion lattice. Each unit cell possesses a single central $s=2$ umbilic, with the six outer $s=1$ umbilics each shared between $3$ unit cells giving a total count of $2+6\times 1/3=4$ umbilics $=1$ Skyrmion per unit cell. \textit{Top Left}: Unit vector field $\mathbf{n}$ in a Skyrmion lattice. \textit{Top Right}: Contour plot of $|\Delta|$ showing the umbilic lines forming a lattice of tubes. \textit{Bottom Right}: Eigenvector field of $\Pi$, illustrating the $+1$ winding around the central umbilics and $-1/2$ winding around the outer umbilics. Note that both are counted with the same sign as the orientation of $\mathbf{n}$ changes from one to another. \textit{Bottom Left}: Umbilics form lines in three dimensions.}
\label{fig:lat}
\end{figure}

One can understand this relation between the profile and index in terms of oriented umbilics. The 1-form $A$ circulates around umbilic lines, orienting them such that the circulation is counter-clockwise. If one orients umbilics in this way, the expression for the index in \eqref{eq:gbc} is rewritten as 
\begin{equation}
 \textrm{index}_\Sigma \; U_i = s_i \textrm{Int}(U_i,\Sigma),
\end{equation} 
where $\textrm{Int}(U_i,\Sigma)$ is the signed number of intersections of $U_i$ with $\Sigma$ and $s_i \in \mathbb{N}$ is the absolute strength of the umbilic $U_i$. Examples illustrating this are shown in Figs.~\ref{fig:vort} and~\ref{fig:lat}; we describe only the latter (the former is essentially identical). Fig.~\ref{fig:lat} shows a Skyrmion lattice, a well-known structure seen in a variety of systems~\cite{rossler06,muhlbauer09,seki12,yu10,yu11,kiselev11,tonomura12,yu12,choi13}. The umbilics form a lattice, with $s=2$ umbilics each surrounded by six $s=1$ umbilics. The eigenvectors of $\Pi$ show the central umbilic having a $+1$ profile, and the six outer umbilics a $-1/2$ profile. All the umbilics have the same orientation induced by $A$, and so their intersection numbers, and thus indices in \eqref{eq:gbc}, are all positive. For the $+1$ profile, $-1/2$ profile and the orientations to all be consistent, it is necessary that $\mathbf{n}$ rotates by $\pi$ when passing from the central umbilic to an outer one.

The curvature of the bundle $\xi$ can be expressed in terms of the director field as 
\begin{equation}
\Omega_{\xi} = - \frac{1}{2} \epsilon_{abc} n^{a} \partial_i n^{b} \partial_j n^{c} dx^i \wedge dx^j .
\end{equation}
This is because each 2-plane of the bundle $\xi$ is explicitly embedded in $\mathbb{R}^3$ (see Fig.~\ref{fig:2plane}) and the expression is just the extrinsic definition of curvature in terms of the (generalised) Gauss map. Equally, it follows from an explicit calculation in local coordinates for any particular texture, say a radial hedgehog. A third way to establish it is through a calculation in the style of the Mermin-Ho relation~\cite{mermin76}, a good account of which has been given by Kamien~\cite{kamien02}. The curvature of $E$ is equal to twice this, $\Omega_E=2\Omega_{\xi}$, since $E$ is a spin 2 subbundle of $\xi^*\otimes\xi$. 
Therefore, when $\Sigma$ is closed, or the boundary term vanishes, we have 
\begin{equation}
\sum_{i} s_i \textrm{Int}(U_i,\Sigma) = \frac{1}{2\pi} \int_{\Sigma} \epsilon_{abc} n^a \partial_i n^b \partial_j n^c dx^i \wedge dx^j = 4q ,
\label{eq:Skyrmion_degree}
\end{equation}
or the sum of the indices of the umbilics piercing $\Sigma$ is equal to four times the Skyrmion number. This establishes a general relationship between Skyrmions and umbilics. This is illustrated in Fig.~\ref{fig:lat} where each unit cell contains one $s=2$ umbilic and six $s=1$ umbilics, each of which are shared between three unit cells, giving a total count of $2+6\times 1/3 = 4$, \textit{i.e.} there is one Skyrmion in each unit cell. This approach can also be extended to the charcterisation of merons, or half-Skyrmions, which can be thought of as a single $s=2$ umbilic, or two $s=1$ umbilics, as in the case of the cholesteric dislocation, shown in Fig.~\ref{fig:vort}. 

This correspondence between Skyrmions and umbilics can be seen as a version of Poincar\'{e} duality, giving a natural way to localise Skyrmions to a set of points or lines. The set of umbilics $\mathcal{U}$ in a three dimensional domain $M$ gives a representative of a cycle in $H_1(M)$. The Poincar\'{e} dual of this is the cocycle $e(E) \in H^2(M)$, the Euler class, which depends only on the topology of the bundle $E$ and hence $\mathbf{n}$, and gives a global constraint on the total number of umbilic lines in the system in terms of the topology of $\mathbf{n}$. As a spin-2 vector bundle, $e(E)=2e(\xi) = 4[\mathbf{n}]$, where $[\mathbf{n}] \in H^2(M)$ is the cocycle represented by $\mathbf{n}$~\cite{note classes}. In the absence of torsion in $H^2(M)$~\cite{note torsion}, there is a canonical representative of the cocycle $4[\mathbf{n}]$ as a 2-form which is given through the Chern-Weil homomorphism by $\textrm{Pf}(-\sfrac{\Omega_E}{ 2 \pi})$, which leads to \eqref{eq:gbc}. 

\begin{figure}
\includegraphics{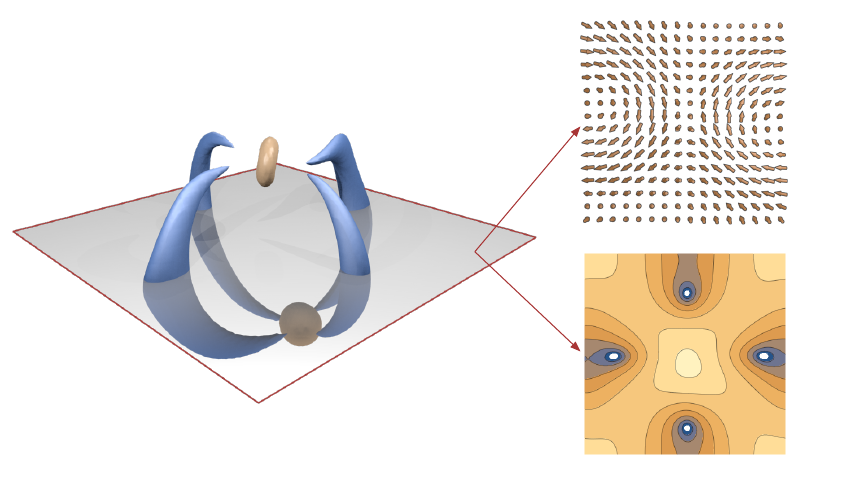}
\caption{Transient monopoles and umbilic lines observed in a simulated quench of the free energy \eqref{eq:f}. {\it Left:} Contours of constant value of $| \Delta |$. The blue lines are umbilics, the beige spheres are monopoles (point defects). Each umbilic has winding $n=1$. They are all oriented similarly, so that the sum of the umbilics entering/leaving the negative/positive point defect is 4, in accordance with \eqref{eq:gbc} and corresponding to the Skyrmion created between the monopoles. {\it Top Right:} Cross-section showing $\mathbf{n}$ on the shaded plane. As a map from the plane to $S^2$, it has degree 1, indicating the presence of a Skyrmion. {\it Bottom Right:} Contour plot showing the magnitude of $\log(| \Delta |)$ on the shaded plane, the umbilics are clearly picked out.}
\label{fig:mono}
\end{figure}

An example of the Euler class as a global constraint on umbilics is given by the case of monopoles, illustrated in Fig.~\ref{fig:mono}. These are point defects in the magnetisation that mediate changes in the number of Skyrmions~\cite{milde13}. In the local neighbourhood of a point defect the topology of $\mathbf{n}$ is described by an integer $q \in \pi_2(S^2) \cong \mathbb{Z}$, the charge of the point defect. Through \eqref{eq:Skyrmion_degree} one finds that a sphere surrounding this point defect must be pierced by umbilics of total index $4q$. In Fig.~\ref{fig:mono} we show a simulation of two $\pm 1$ point defects, or monopoles, in a chiral ferromagnet. There are four umbilics, each of generic type, ending on the point defects in a manner reminiscent of Dirac strings~\cite{dirac31}. On a slice in the midplane of the cell $\mathbf{n}$ contains a Skyrmion, illustrating the well-known creation and annihilation of Skyrmions in terms of point defects, observed experimentally in \cite{milde13}.

\section{Examples of Line Fields: Blue Phases}

So far we have confined our analysis to those systems where the orientational order is represented by a vector. In liquid crystalline phases the orientational order is instead line-like. Perhaps the most noticeable feature of this is the presence of line defects, called disclinations, in liquid crystals~\cite{mermin79,alexander12}, which make manifest the non-orientability and are absent in materials with vector order. However, they can arise even in vector ordered materials if the vector couples to an internal degree of freedom to give nematic-like symmetry; examples include thin films of $^3$He~\cite{korshunov86} and spin 1 polar condensates~\cite{james11}. 

As the director is non-singular along umbilics, the effect of the non-orientability is primarily restricted to their global identification. Indeed, since umbilics can be identified from the local structure of the director gradients, they have the same local description for line fields as they do for vector fields. One need only choose an orientation for the line field, which may always be done locally, {\it i.e.} in regions that do not contain disclination lines. There is even the same description for their global character in that the umbilics still furnish a representative of the Poincar\'e dual to twice the Euler class of the orthogonal 2-plane bundle $\xi$. Interestingly, for line fields with knotted disclination lines the Euler class of the orthogonal 2-plane bundle is usually torsion~\cite{machon14}, which the usual Chern-Weil theory of curvature forms is insensitive to. (The curvature form represents only the integral part of the cohomology~\cite{MilnorStasheff,BottTu}.) The integral in \eqref{eq:Skyrmion_degree} thus does not provide any information in such cases. However, as the umbilic lines are Poincar\'e dual to twice the Euler class of $\xi$ (and $4[{\bf n}]$) they continue to capture the topology of the orientational order, even when it is torsion, modulo elements of order 4.

As discussed earlier, and shown in Fig.\ \ref{fig:vort}, one can identify umbilics with $\lambda$ lines in cholesterics. In this way, \eqref{eq:gbc} can be seen as a global constraint on the $\lambda$ lines in a cholesteric, as determined by the topology of $\mathbf{n}$, with additional subtlety associated to elements in $H^2(M)$ of order two and four. We anticipate that the elements of order 4 are associated with $\tau$ lines. For the elements of order 2 the orthogonal 2-plane bundle $\xi$ admits a global section and we anticipate that these are associated with distinct ``cholesteric ground states''~\cite{machon15}.

Rather than present a detailed account of the analysis of line fields, we focus on simple illustrations and a practical means for identifying umbilics globally, using one particular physical system -- the cholesteric blue phases. These remarkable materials are characterised by periodic orientational order corresponding to crystalline space groups, while remaining three-dimensional fluids. Two distinct cubic structures, BPI with space group $O^{8-}(I4_132)$ and BPII with space group $O^2(P4_232)$, are observed on varying temperature or the amount of chiral dopant~\cite{meiboom81,grebel83,wright89}, while several other structures can be induced by applied electric field effects~\cite{pieranski85,pieranski87,hornreich90} or confinement~\cite{fukuda10b}, and there is also an amorphous phase, known as BPIII~\cite{henrich11}. Their structure arises because the local optimal configuration for the chiral free energy is one of double twist (illustrated in Fig.~\ref{fig:vort} and \ref{fig:2plane}), but this is only locally favourable and the structure becomes energetically unstable if it extends too far. This leads to a frustrated arrangement where locally preferred regions of double twist form periodic arrays interwoven by a network of disclination lines~\cite{meiboom81,wright89}. The blue phases have a long-standing association with Skyrmions in chiral magnets, with which they may be considered analogous. This analogy continues to stimulate considerable interchange of ideas between the two fields~\cite{binz06,fukuda10,hamann11,ackerman14,leonov14}.

It is usual to describe the order in liquid crystals, especially in numerical simulations, using a tensor order parameter, called the Q-tensor, that captures the non-orientability of the director field~\cite{deGennesProst}. The Q-tensor is proportional to the deviatoric part of the dielectric tensor or the second moment of the molecular orientational distribution function. The director field can be recovered from it as the eigenvector associated to the largest eigenvalue. In typical nematic materials, the Q-tensor is uniaxial and we have ${\bf Q}\propto {\bf n}\otimes{\bf n}-\tfrac{1}{3}\mathbb{I}$. Our first objective is to describe how umbilics may be identified from the Q-tensor.  

Orientational order continues to define a splitting of the tangent bundle $T\mathbb{R}^3\cong L_{{\bf n}}\oplus\xi$ whether it is orientable or not, except that the line bundle $L_{{\bf n}}$ is non-trivial in the non-orientable case, with the consequence that the director ${\bf n}$ is not globally defined (as a vector). Rather, the projector ${\bf n}\otimes{\bf n}$ provides a section of $L_{{\bf n}}\otimes L_{{\bf n}}$. Similarly, the shape operator has to be modified so as to be presented as a section of the twisted bundle $\xi^*\otimes\xi\otimes L_{{\bf n}}$~\cite{machon15}. However, interestingly, the chirality pseudotensor, and its deviatoric part $\Pi$ \eqref{eq:def_Pi}, continue to be globally well-defined in the same form even when the director is non-orientable, so that $\Pi$ admits an immediate extension to a `Q-tensor version'. We shall use this to identify umbilics in line fields. 

As $\xi^*\otimes\xi$ is a subbundle of $T\mathbb{R}^{3\,*}\otimes T\mathbb{R}^3$ the linear transformation $\Pi$, equation~\eqref{eq:def_Pi}, can be embedded in a $3\!\times\!3$ matrix and expressed, in a Cartesian basis, as 
\begin{equation}
\begin{split}
\Pi_{ij} & = \frac{1}{4} \epsilon_{ilk} \Bigl[ n_l \partial_k n_j + n_l \partial_j n_k - n_j n_l n_m \partial_m n_k \Bigr] \\
& \quad + \frac{1}{4} \epsilon_{jlk} \Bigl[ n_l \partial_k n_i + n_l \partial_i n_k - n_i n_l n_m \partial_m n_k \Bigr] .
\end{split}
\end{equation}
The issue at hand is what a suitable replacement should be if we express the orientational order using a Q-tensor rather than the director field. The usual approach to obtaining such a replacement is to adopt algebraic methods rather than geometric ones~\cite{atiyah_quote}; one constructs a list of potential expressions and selects on the basis of how they reduce when the Q-tensor is written in terms of the director field. Following this procedure, we find that the expression 
\begin{equation}
\begin{split}
\tilde{\Pi}_{ij} & = \frac{1}{4} \epsilon_{ilk} \Bigl[ 2 Q_{lm} \partial_k Q_{jm} + Q_{jm} \partial_k Q_{lm} \\
& \qquad \qquad + Q_{lm} \partial_j Q_{km} - Q_{jl} \partial_m Q_{km} \Bigr] \\
& \quad + \frac{1}{4} \epsilon_{jlk} \Bigl[ 2 Q_{lm} \partial_k Q_{im} + Q_{im} \partial_k Q_{lm} \\
& \qquad \qquad+ Q_{lm} \partial_i Q_{km} - Q_{il} \partial_m Q_{km} \Bigr] ,
\end{split}
\end{equation}
reduces to $\Pi_{ij}$ when the Q-tensor is uniaxial, with constant magnitude, $Q_{ij}\propto n_in_j-\tfrac{1}{3}\delta_{ij}$. We subtract its trace and adopt it as an appropriate Q-tensor analogue of $\Pi$. In simulations, the umbilic lines can then be identified using isosurfaces where the norm of $\tilde{\Pi}$ drops below a threshold value. 

\begin{figure}
\includegraphics{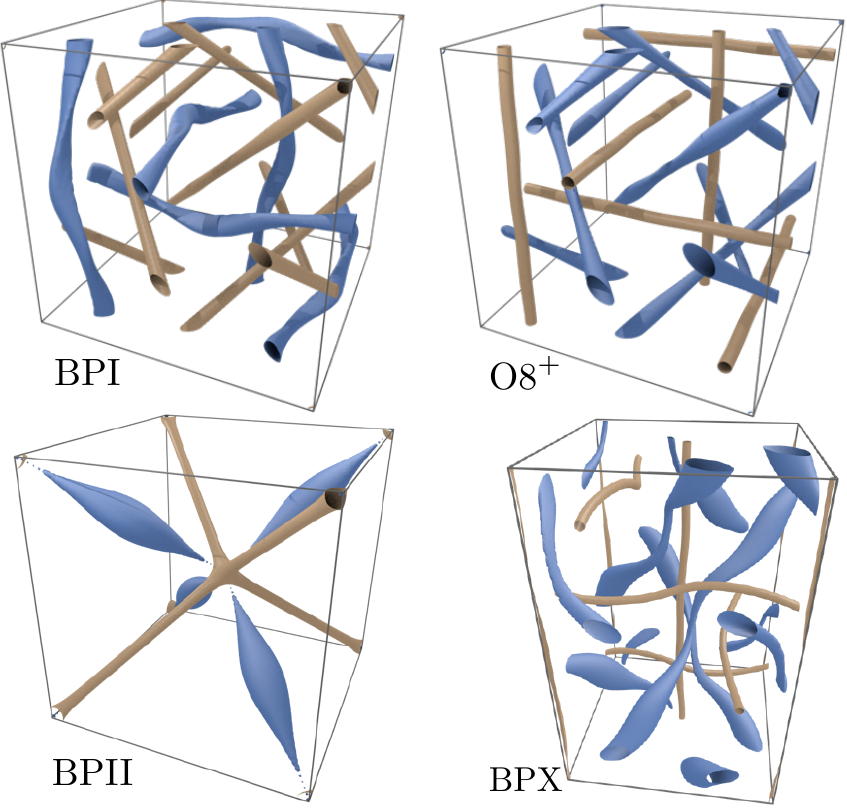}
\caption{Umbilic lines in the blue phases. The umbilics are indicated by blue isosurfaces, while the network of disclination lines is shown by milk chocolate isosurfaces. The textures BPI and BPII are the cubic structures observed experimentally. $O8^{+}$ is a cubic structure with the same space group as BPI but an interchange between disclination lines and umbilics. BPX is a tetragonal texture observed in an applied electric field. The arrangement of disclinations and umbilics is the same as in $O^{8+}$, although the symmetry is different.}
\label{fig:bp}
\end{figure}

Illustrations are given in Fig.~\ref{fig:bp} for some standard blue phase textures. The umbilics coincide with structural motifs that have been identified previously, either through symmetry considerations~\cite{hornreich88} or by taking a singular value decomposition of the Q-tensor~\cite{copar13}. These motifs are the axes of double twist cylinders~\cite{meiboom81,wright89}. Two remarks are in order concerning this. First, the pattern of umbilic lines in BPII (see Fig.~\ref{fig:bp}) does not correspond to the array of double twist cylinders traditionally depicted in the literature. In the traditional depiction the double twist cylinders are taken along the 2-fold axes lying in the faces of the conventional unit cell and, indeed, inspection by eye reveals these lines to have the same general structure as model double twist cylinders~\cite{meiboom81} (see Fig.~\ref{fig:vort}). However, in a Q-tensor description of BPII these are not degeneracies where the local structure is axisymmetric, as in the standard idealised picture of a double twist cylinder; rather they are 2-fold lines where the Q-tensor is maximally biaxial~\cite{GPA_thesis,ravnik11,copar13}. Second, it is perhaps worth emphasising that the umbilics identified in all other blue phase textures also differ from the standard idealised picture of a double twist cylinder. This is because the idealised axisymmetry of the latter constrains it to be a degenerate winding $+2$ umbilic, whereas those that are found in periodic textures are generic umbilics, typically with a $-1/2$ profile for the eigenvectors of $\tilde{\Pi}$, {\it i.e.} a star umbilic.  

We have given here only cursory illustrations for periodic blue phases with crystalline space groups. It would be of interest to also study the structure of the amorphous BPIII in terms of umbilics, as well as their behaviour in transitions and rheology. The Q-tensor version of the linear transformation $\Pi$ that we introduced here was motivated on algebraic grounds rather than the geometrical considerations that the rest of our work is founded upon. It is clear that geometrical considerations can be developed for traceless symmetric tensors, parallel to our description of orientational order. It would be interesting to do so and compare the outcome with the algebraic construction we have given. 

In the remainder of this paper we return to the setting of vector orientational order.

\section{Local Profiles of Umbilic lines}

Generic umbilic lines carry a canonical orientation through the vector $B$, defined as
\begin{equation}
B^i= \langle \Pi,\partial^i \Delta \rangle = \langle \Delta,\Delta \rangle \delta^{ij} A_j,
\end{equation}
which assigns a tangent vector to the umbilic, provided it is generic, as follows. If, in a local trivialisation, $\Delta=s_1 \mathbf{e}_1+s_2 \mathbf{e}_2$ then an umbilic is defined as the intersection of the two surfaces $s_1=0$ and $s_2=0$ and so its tangent vector will be $\nabla s_1 \times \nabla s_2$. Note that while the values of $s_1$ and $s_2$ depend on the choice of basis, the tangent vector does not. In this local form we can write
\begin{equation}
B = s_1 \nabla s_2 - s_2  \nabla s_1 + 2(s_1^2+s_2^2)\omega ,
\end{equation}
so that on an umbilic we have $\nabla \times B = 2 \nabla s_1 \times \nabla s_2$ and hence the vector $\nabla \times B$ points along umbilic lines, canonically orienting them. As indicated in Fig. \ref{fig:anat} this orientation is compatible with that induced by the circulation in $A$ previously discussed. As alluded to in Figs.~\ref{fig:vort} and \ref{fig:lat}, the discrepancy between the type of profile and index of an umbilic depends on the relative orientation between $\mathbf{n}$ and the umbilic line. To measure whether the eigenvectors of a generic umbilic have winding $\pm 1/2$ one should integrate $A$ along a contour $\gamma$ chosen such that its orientation matches the orientation imparted to $\xi$ by $\mathbf{n}$. One can then show that the local profile is $\pm1/2$ depending on 
\begin{equation}
\textrm{Sgn}(\mathbf{n} \cdot \nabla \times B) ,
\label{eq:wind}
\end{equation}
the analysis being identical to that given by Berry for C lines in electromagnetic fields~\cite{berry04b}. 

This quantity $\mathbf{n} \cdot \nabla \times B$ can be used to further characterise the local structure of umbilic lines. Along an umbilic $\Delta$ vanishes and generically we can expect it to vanish linearly. Its local profile is then given by the form of the linear order terms in $\Delta$ in the immediate vicinity of the umbilic. Equivalently, generically we can expect the gradients of $\Delta$ to be non-zero on the umbilic and the local profile is then given by the structure of $\nabla\Delta$ evaluated on the umbilic. As usual, we split the gradients into the components parallel to ${\bf n}$ and those perpendicular to it, and consider only the latter. The non-zero part of these transverse gradients, evaluated on the umbilics, takes values in $\xi^*\otimes E$, so it is this part that we need to focus on. A section of this bundle characterising the local profile of the umbilic can be constructed from $\nabla\Delta$ as follows. If $\mathbb{I}_E$ is the projection onto $E$ in $T\mathbb{R}^{3\,*} \otimes T \mathbb{R}^3$ then we define the differential operator $\nabla^\xi$ as $\nabla^\xi \Delta = \mathbb{I}_\xi \nabla \Delta \mathbb{I}_E$. By construction $\nabla^\xi \Delta$ is a section of the bundle $\xi^* \otimes E$. 
Under the action of $SO(2)$ the bundle $\xi^*\otimes E$ splits into two rank 2 subbundles, $F^{+}\oplus F^{-}$, where $F^+$ is the spin 1 bundle corresponding to $+1/2$ profiles and $F^-$ is the spin 3 bundle corresponding to $-1/2$ profiles, and we decompose $\nabla^{\xi}\Delta$ according to this splitting as  
\begin{equation}
\nabla^{\xi} \Delta = \nabla^{\xi} \Delta^{+} + \nabla^\xi \Delta^{-} .
\end{equation}
A short calculation shows that $\mathbf{n} \cdot \nabla \times B=|\nabla^\xi \Delta^+|^2-|\nabla^\xi \Delta^-|^2$ so that if $|\nabla^{\xi}\Delta^{+}|>|\nabla^{\xi}\Delta^{-}|$ the umbilic has a local $+1/2$ profile, while if the converse holds the profile is of type $-1/2$. 

This analysis allows us to give a description of the space of local profiles for a generic umbilic. Since $\nabla^{\xi}\Delta|_{{\cal U}}$ is non-zero for a generic umbilic, but otherwise arbitrary, the space of local profiles has the homotopy type of the three-sphere. However, there is more structure to it than that. The equality $|\nabla^{\xi}\Delta^{+}|=|\nabla^{\xi}\Delta^{-}|$ defines a Clifford torus separating the three-sphere into two solid tori corresponding to the different profiles. A natural constraint on an umbilic loop is for its profile to be of constant type, \textit{i.e.} for $\mathbf{n} \cdot \nabla \times B$ to be of constant sign or the profile to stay within a single solid torus. We call an umbilic loop which satisfies this constraint transverse, as its tangent vector is always transverse to the planes of $\xi$. This is not an unphysical constraint, transverse umbilic lines are commonly found in various chiral structures -- typical $\lambda$ lines in cholesterics have a local structure in which $\mathbf{n}$ is either parallel or antiparallel with the tangent vector to an umbilic. Figure \ref{fig:toron} shows the transverse umbilic loop found in toron excitations in frustrated cholesterics~\cite{smalyukh10,chen13}. 

The condition of an umbilic being transverse therefore translates as one of the two profiles having a higher `weight' along the entire length of the umbilic. If this is the case then we have $|\nabla^\xi \Delta^\pm|^2 \neq 0$ with the sign depending on whether the umbilic is positively or negatively transverse. Suppose, for concreteness, that the umbilic is positively transverse. Then, in a local trivialisation, we will have $\nabla^\xi \Delta^+ = t_1 \mathbf{f}^+_1+t_2 \mathbf{f}^+_2$, with $t_1^2+t_2^2 \neq 0$, and the space of positively transverse umbilics has the homotopy type of a circle. The variation of the profile around a closed, transverse umbilic loop confers another integer winding number. This number depends on how the basis vectors of the trivialisation of $F^+$ wind around the umbilic. The ambiguity can be removed by demanding  that each of the basis vectors in the trivialisation does not link with the umbilic. As we will show in the next section, this winding number can be interpreted as twice a self-linking number for the umbilic loop, and is related to a relative Euler class of $E$.

\section{Umbilic Loops} 

The umbilics of three-dimensional orientationally ordered materials are extended line-like objects, however most of the properties we have described so far can be associated to isolated points along these one-dimensional curves. For example the points of transverse intersection with an appropriate surface (or slice through the material) count numbers of Skyrmions, merons, or $\lambda$ defects. It is clear, however, that these locally measurable objects do not represent all the information encoded by umbilic lines; if one has a closed umbilic loop then, as discussed above, the local proprties at each point must stitch together consistently along the loop conveying both geometric and topological information about the orientational order. 

\begin{figure}
\includegraphics{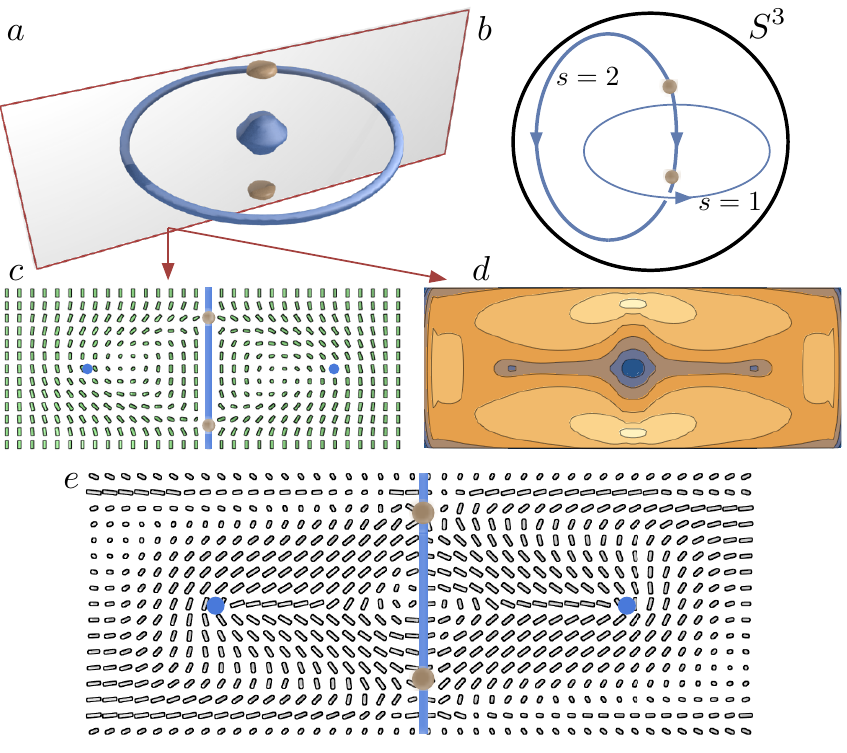}
\caption{Umbilic lines in a toron. (a) Simulation results showing point defects and characteristic umbilic loop in a toron. The umbilic loop is transverse and has zero self-linking number. The point defects have opposite charge, due to lattice effects the simulation does not resolve the umbilic lines connecting them, though they are clear in the topology of the eigenvector field. (b) Diagram illustrating the umbilic structure of the toron in $S^3$ when the boundary of the cell is collapsed to a point, note that the umbilic loops connecting the point defects each have strength $s=2$, for a total of 4, as required by \eqref{eq:gbc}. (c) Simulation results showing director field for the toron. (d) Contour plot showing the value of $|\Delta|$. (e) Eigenvector field of $\Pi$. Note that the umbilic loop is generic and has a monstar profile.}
\label{fig:toron}
\end{figure}

Umbilic loops have been observed in experimental systems~\cite{ackerman14,chen13,flossmann08}, and a motivating set of examples are the so-called torons~\cite{smalyukh10}, illustrated in Fig. \ref{fig:toron}. These soliton structures can be generated in a frustrated cholesteric cell, with the director held vertical along the top and bottom. As shown in Fig. \ref{fig:toron}(a), they display a characterisitc umbilic loop, in the form of a circle sitting in the midplane of the cell. This loop is of generic type, with the eigenvectors of $\Pi$ having a monstar profile. 

As with local properties of umbilic lines, much of the structure of loops is encoded in the 1-form $A$. Previously we have obtained local information about umbilics by integrating $A$ around a meridian. The first step to understanding their global structure is to integrate $A$ along a longitude. Since $A$ is singular on an umbilic, we must define this integral as a limit. If $U_i$ is an umbilic, then let $U_i^\prime$ be a curve close to $U_i$ that has zero linking number with it. The integral of $A$ over $U_i^\prime$ is well defined and if all the umbilics in the system are closed loops we can take the limit $U_i^\prime \to U_i$ to obtain
\begin{equation}
\int_{U_i} A  =  2 \pi \sum_j s_j \textrm{Lk}(U_i,U_j) + \int_{\Sigma} \Omega_E ,
\label{eq:relchern}
\end{equation}
where $\Sigma$ is an orientable surface bounded by $U_i$. The requirement that $U_i^\prime$ have zero linking with $U_i$ ensures that the limit is well defined but in general the result depends upon this choice of framing. The quantity $\sum_j s_j \textrm{Lk}(U_i,U_j)=e(U_i)$ in \eqref{eq:relchern} is an integer associated to the umbilic line, the relative Euler class, and counts the number of umbilics linking $U_i$. There are two natural questions to ask of this quantity. The first is whether it corresponds to any local geometric structure of the umbilic line; the second is whether it relates to a global property of $\mathbf{n}$. On both counts the answer is yes. Geometrically \eqref{eq:relchern} is related to both a self-linking number and the winding number defined in the previous section. Globally we will show that it is related to the Hopf invariant of $\mathbf{n}$. 

\begin{figure}
\begin{center}
\includegraphics{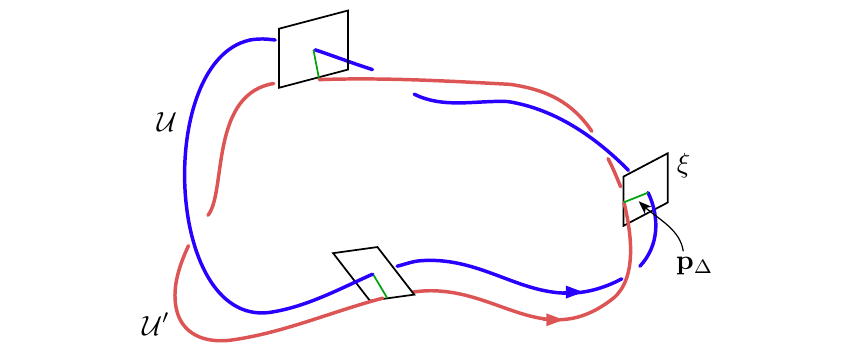}
\end{center}
\caption{The self-linking number for transverse umbilic loops. The red curve $U^\prime$ is formed by pushing $U$ along an eigenvector field $\mathbf{p}_\Delta$ of $\Delta$. The self-linking number is then computed as the linking number of $U$ with $U^\prime$. The umbilic must be transverse to ensure that $U^\prime$ does not cross $U$. Because the eigenvectors are degenerate on the umbilic, one evaluates $\mathbf{p}_\Delta$ on a line close to $U$, $\hat{U}$, that has zero self-linking number with $U$. Such a line can be found by drawing an orientable surface whose boundary is $U$ and pushing $U$ into that surface. Note that because eigenvectors are line fields, the self-linking number can be a half-integer, corresponding to $\mathbf{p}$ being non-orientable along the longitude of $U$.}
\label{fig:sl}
\end{figure}

To see this suppose $U$ is a transverse umbilic loop, then let $\hat{U}$ be a loop close to $U$ of zero linking number with $U$. Evaluate one of the eigenvectors of $\Delta$, $\mathbf{p}^+_\Delta$ say, along $\hat{U}$ and then push $U$ along this eigenvector field to form a new loop, $U^\prime$. The linking number $\textrm{Lk}(U,U^\prime)$ is then taken to define the self-linking number of the transverse umbilic, $\textrm{Sl}(U)$. This construction is shown in Fig.\ \ref{fig:sl}. Note that since $\mathbf{p}^+_\Delta$ is a line field, $\textrm{Sl}(U)$ may be a half-integer corresponding to $\mathbf{p}$ being non-orientable along the longitude of $U$.

With this notion in hand the correspondence between the self-linking number and the winding number coming from the variation of the local profile defined in the previous section can be established. Choose a trivialisation $\{{\bf d}_1,{\bf d}_2\}$ of $\xi$ on a neighbourhood of $U$ such that the linking number of $U$ pushed off along $\mathbf{d}_1$ (say) is zero then the winding of $\nabla^\xi \Delta^+$ with respect to the chosen trivialisation will give twice the self-linking number. Note that the geometric definition of the self-linking number did not depend on the zero of $\Delta$ being of linear order. The relationship between between the self-linking number and the relative Euler class of a transverse umbilic is then~\cite{note sl}
\begin{equation}
\textrm{Sl}(U) = \overline{\textrm{Sl}}(U)+ \frac{1}{2} e(U).
\label{eq:selflinking}
\end{equation}
There is a new term in this equation, $ \overline{\textrm{Sl}}(U)$, the transverse self-linking number of $U$. $ \overline{\textrm{Sl}}(U)$ is a geometric quantity, preserved under deformations that keep $U$ transverse. A similar notion is studied in the context of contact topology~\cite{Geiges} where it provides useful geometric information, but we do not know of a general discussion in the case of arbitrary $\mathbf{n}$. 

\begin{figure}
\includegraphics{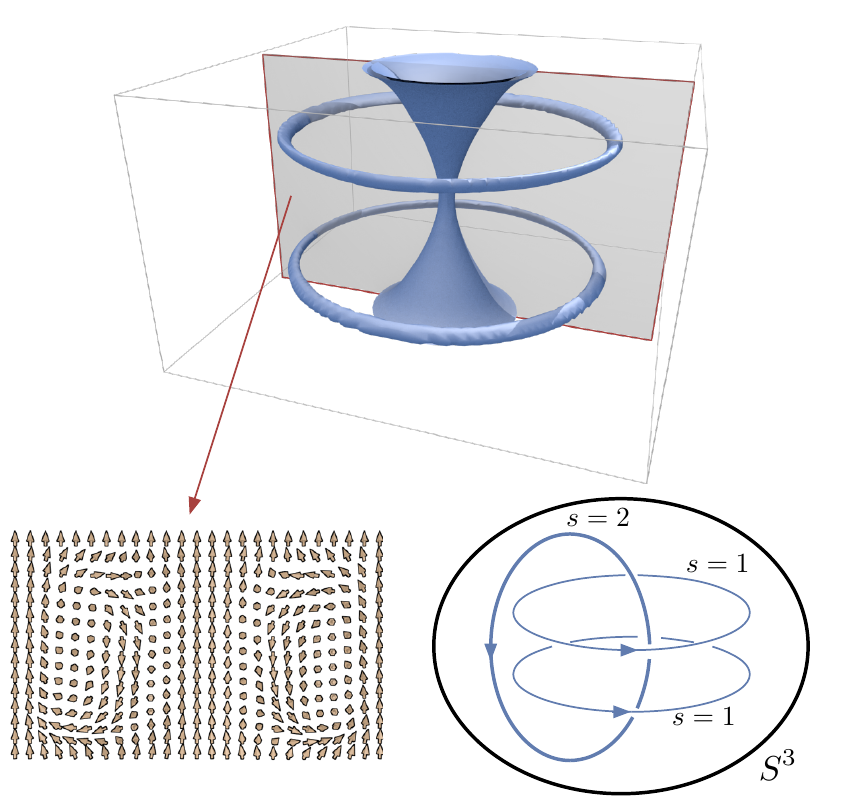}
\caption{Umbilic lines in the Hopf fibration. \textit{Top}: Umbilic lines given as isosurfaces of $|\Delta|$ for a simulated Hopf fibration. The two lines in the mid plane of the cell are both generic, with the same orientation. For symmetry reasons the central umbilic is not generic and has $s=2$. On the boundary of the cell $\mathbf{n}$ is uniform, which means that the Hopf invariant is well-defined. Note that the system is totally umbilic on the boundary. If you compactify the domain by collapsing the boundary to a point, then one can interpret the central vertical umbilic line as passing through this point. \textit{Bottom Left}: director field for the Hopf texture. \textit{Bottom Right}: Schematic of the relationship between umbilic lines in $S^3$. Two generic $s=1$ umbilic lines each link with one $s=2$ umbilic line that passes through the point at infinity.}
\label{fig:hopf}
\end{figure}

We now turn to the relationship between the linking of umbilics and global properties of $\mathbf{n}$. Associated to the 1-form $A$ is the Chern-Simons 3-form, $A \wedge d A$, and it is natural to ask whether the integral over the complement of $\mathcal{U}$,
\begin{equation}
\int A \wedge dA,
\label{eq:ch3hopfint}
\end{equation}
gives some information about umbilic lines and $\mathbf{n}$. This integral can be compared to the similar helicity integral for a fluid flow and the Abelian Chern-Simons action. Both these quantities are related to linking numbers, of vortex lines in the case of helicity~\cite{moffatt69,arnold86,moffatt92} and Wilson loops in Chern-Simons theory~\cite{witten89,polyakov88} and, as we will show, it is no different for umbilics, \eqref{eq:ch3hopfint} connects the linking numbers of the umbilic loops with the Hopf invariant of $\mathbf{n}$. To proceed we will assume that the material domain is $S^3$, with no defects and that the zero set of $\Delta$ is one-dimensional. By a compactification, this assumption that the domain is $S^3$ is equivalent to examining a system in $\mathbb{R}^3$ for which $\lim_{x \to \infty} \mathbf{n}(x) = \mathbf{n}_0$, a constant. An example of such a system is illustrated by the Hopf fibration, which has been experimentally realised in frustrated cholesterics~\cite{chen13}, illustrated in Fig.\ \ref{fig:hopf}, and the aformentioned toron. In both these systems  $\mathbf{n}$ is constant on the boundary of the cell. Because of this $\Delta =0 $ on the boundary and one should consider these systems as having an umbilic line passing through the point `at infinity', which in Figs.~\ref{fig:toron} and \ref{fig:hopf} is just the boundary of the system. 

Topologically the configuration in Fig.\ \ref{fig:hopf} is unusual, it is non-singular but not homotopically equivalent to a constant. The difference is that it has a non-zero Hopf invariant. This integer invariant distinguishes topologically distinct non-singular textures in $\mathbb{R}^3$ which are constrained to tend to a constant at infinity and is typically visualised in terms of linking of pre-images of two orientations using the Pontryagin-Thom construction~\cite{chen13}. In our setting however one can give a slight re-interpretation of the famous Whitehead integral formula~\cite{whitehead47}, similar to that given by Arnol'd~\cite{arnold86} to show that the Hopf invariant is also the Chern-Simons invariant of the bundle $\xi$. In our interpretation, if one takes $\omega$ to be a connection form for $\xi$, then $d \omega$ is the curvature of $\xi$ and the Chern-Simons integral
\begin{equation}
\frac{1}{(4 \pi)^2}\int_{S^3} \omega \wedge d \omega
\end{equation}
computes the Hopf invariant of $\mathbf{n}$. 

Now we turn back to our integral \eqref{eq:ch3hopfint}. As discussed, we assume the material domain is $S^3$ (or $\mathbb{R}^3$ with uniform boundary conditions). This means that we can give a global trivialisation of $\xi$ with connection $\omega$ and our 1-form $A$ can be written as $A = d\theta + 2 \omega$. This allows us to write
\begin{equation}
\int_{S^3-\mathcal{U}} A \wedge dA = 4 \int_{S^3-\mathcal{U}} \omega \wedge d \omega + 2\int_{S^3-\mathcal{U}} d \theta \wedge d \omega
\end{equation}
or, since $\mathcal{U}$ has measure zero
\begin{equation}
\int_{S^3-\mathcal{U}} A \wedge dA = 4 (4 \pi)^2H + 2\int_{S^3-\mathcal{U}} d \theta \wedge d \omega
\end{equation}
where $H$ is the Hopf invariant of $\mathbf{n}$. Now we need to evaluate the integral of $d \theta \wedge d \omega$ over $S^3-\mathcal{U}$. We will first do a simpler integral, over $S^3-N(\mathcal{U})$, where $N(\mathcal{U})$ is a neighbourhood of the umbilic lines. By Stokes' theorem this can be reduced to an integral over the boundary of $N(\mathcal{U})$ as
\begin{equation}
\int_{S^3-N(\mathcal{U})} d \theta \wedge d \omega = \sum_i \int_{\partial N(U_i)} \omega \wedge d \theta.
\end{equation}
The boundary of the neighbourhood of each umbilic line is a torus and because our domain is $S^3$ we can choose a preferred longitude for each torus that has zero linking with the umbilic at its centre. This allows us to specify a decomposition of each boundary component into $S_M^1 \times S_L^1$. If we choose $N(\mathcal{U})$ small enough, then along each meridian of the torus $\omega$ will be approximately constant and equal to its value on the umbilic line. The integral over each boundary component can then be written as:
\begin{equation}
\sum_i \int_{\partial N(U_i)} \omega \wedge d \theta =  \sum_i\int_{S_L^1} \omega^\ast \left ( \int_{S_M^1} d \theta^\ast \right )
\label{eq:minus}
\end{equation}
where we have pulled the forms back to each $S^1$. 

The integral of $d \theta$ is trivial, it evaluates to $s_i 2 \pi$, where $s_i$ is the strength of the $i$\textsuperscript{th} umbilic. Finally, we take a limit as $N(\mathcal{U}) \to \mathcal{U}$ and obtain a sum of Wilson loop integrals
\begin{equation}
\lim_{N(\mathcal{U}) \to \mathcal{U}} \sum_i \int_{\partial N(U_i)} \omega \wedge d \theta = 2\pi \sum_i   s_i \int_{U_i} \omega.
\end{equation}
Now from the Gauss-Bonnet-Chern theorem \eqref{eq:gbc} we can relate intergals of $\omega$ along umbilics to integrals of $A$. Putting this together with the above equations we obtain
\begin{align}
\frac{1}{(4\pi)^2} \int_{S^3-\mathcal{U}} A \wedge dA -  \frac{1}{8\pi}\sum_{i}s_i \int_{U_i} A \notag \\ =  4  H - \frac{1}{4} \sum_{i j} s_i s_j \textrm{Lk}(U_i,U_j)
\label{eq:relation}
\end{align}
This equation relates analytic data, the integral of $A \wedge dA$ and Wilson loop integrals, to topological data, namely the linking number of the umbilic lines and the Hopf invariant. We hope that this relationship will allow for an interpretation of the creation and destruction of Hopf solitons in terms of umbilic lines and lead to connections between the dynamics of umbilic lines and of vortex lines in fluids. It is also tempting to speculate on the use umbilic loops as generators in a Floer theory of orientational order.

\acknowledgements{
We would like to thank Mark Dennis, Randy Kamien, Dan Beller, Simon \v{C}opar, Ricardo Mosna, Daniel Sussman, and Antoine R\'emond-Tiedrez for useful and invigourating discussions. This work was partially supported by the UK EPSRC through Grant No. A.MACX.0002. T Machon also partially supported by a University of Warwick Chancellor's International Scholarship.
}


\begin{thebibliography}{99}
\bibitem{note terminology} We use the word `Skyrmion' in the context of textures observed in chiral magnets, rather than in its original context of a non-linear field theory of pions. This duplicity seems to be sufficiently well-established, and widely known, that the potential for confusion is minimal.
\bibitem{note classes} If $\mathbf{n}$ is considered as a map into $S^2$, then homotopy classes are given by an element in $H^2(M)$ with additional data associated to the Hopf invariant.
\bibitem{note sl} To see that \eqref{eq:selflinking} is correct it is enough to show that increasing $e(U)$ by 1 increase $\textrm{Sl}(U)$ by $1/2$ coupled with standard arguments from contact topology~\cite{Geiges}.
\end{thebibliography}
\end{document}